\documentclass[12pt,oneside]{article}
\usepackage{epsfig, cite}
\usepackage{color}
\usepackage{ifthen}
\usepackage{graphicx}

\def\p{\partial}

\def\half{{1\over 2}}
\def\({\left(}
\def\){\right)}
\def\[{\left[}
\def\]{\right]}

\def\e{\begin{equation}}
\def\q{\end{equation}}
\def\m{\begin{eqnarray}}
\def\n{\end{eqnarray}}

\begin{document}
\thispagestyle{empty} \setcounter{page}{0}

\vspace{2cm}

\begin{center}
{\huge Eternal Chaotic Inflation is Prohibited by Weak Gravity
Conjecture}

\vspace{1.4cm}

Qing-Guo Huang$^1$, Miao Li$^{2,3}$ and Yi Wang$^{2,3}$

\vspace{3mm}

{\em $^1$ School of physics, Korea Institute for
Advanced Study,} \\
{\em 207-43, Cheongryangri-Dong, Dongdaemun-Gu, }\\ {\em Seoul
130-722, Korea}

\vspace{2mm}

{\em $^2$ Interdisciplinary Center for Theoretical Study,} \\
{\em University of Science and Technology of China,} \\ {\em Hefei, Anhui 230026, P.R. China}\\

\vspace{2mm}

{\em $^3$ Institute of Theoretical Physics, Chinese Academia Sinica,} \\
{\em PO Box 2735, Beijing 100080, P.R. China}

\end{center}

\vspace{-.1cm}

\centerline{{\tt huangqg@kias.re.kr}}

\centerline{{\tt mli@itp.ac.cn}}

\centerline{{\tt wangyi@itp.ac.cn}}

\vspace{1cm} \centerline{ABSTRACT}
\begin{quote}
\vspace{.5cm}

We investigate whether the eternal chaotic inflation can be achieved
when the weak gravity conjecture is taken into account. We show that
even the assisted chaotic inflation with potential $\lambda\phi^4$
or $m^2\phi^2$ can not be eternal. The effective field theory
description for the inflaton field breaks down before inflation
reaches the eternal regime. We also find that the total number of
e-folds is still bounded by the inflationary entropy for the
assisted inflation.

\end{quote}
\baselineskip18pt

\noindent

\vspace{5mm}

\newpage

Inflation \cite{Guth:1980zm,Linde:1981mu,Albrecht:1982wi} is the
leading theoretical paradigm for understanding the early universe
and the origin of the primordial perturbations. In the new version
of inflation, inflaton may begin either in the false vacuum, or in
an unstable state at the top of the effective potential. Then the
inflaton field slowly rolls down to the minimum of its effective
potential. This picture relies on an application of low-energy
effective field theory to inflation.

An interesting phenomenon in inflationary scenario is eternal
inflation. There are actually two kinds of eternal inflation. One is
driven by the quasi-exponential expansion of the meta-stable vacua
themselves. In our paper, we focus on the other version called
slow-roll eternal inflation \cite{Vilenkin:1983xq,Linde:1986fd}. In
this scenario, during the period of inflation, the evolution of the
inflaton field $\phi$ is influenced by quantum fluctuations, which
can be pictured as a random walk of the field with a step $\delta
\phi \sim H/2\pi$ on a horizon scale (Hubble scale $H^{-1}$) per
Hubble time $\Delta t \sim H^{-1}$ \cite{Linde:2005ht}. During the
same epoch, the variation of the classical homogeneous inflaton
field rolling down its potential is $\Delta \phi \sim |\dot \phi|
\cdot H^{-1}$. If the classical evolution is smaller than the
quantum fluctuations, the role played by fluctuations becomes
significant and the inflaton field climbs up its potential rather
than rolls down its potential in some spacetime regions. This leads
to an eternal process of self-reproduction of the universe. See the
recent reviews in \cite{Guth:2007ng,Linde:2007fr}.

Since superstring theory lives in ten dimensions, the six spatial
dimensions must be compactified in order to understand the
four-dimensional physics in the observed universe. Recent
development on the flux compactification
\cite{Giddings:2001yu,Kachru:2003aw,Douglas:2006es} suggests a vast
number of at least semi-classically self-consistent meta-stable
vacua. However not all of the meta-stable vacua are really
self-consistent. The vast series of semi-classically consistent
effective field theories which are actually inconsistent are called
{\em swampland} \cite{Vafa:2005ui}. The series of really
self-consistent vacua, called {\em landscape}, is surrounded by the
swampland. Usually the Planck scale $M_p$ is taken as a naturally UV
cutoff for the effective field theory. But the effective field
theory can break down even in the region with low curvature. In
\cite{ArkaniHamed:2006dz}, the authors suggest a new criterion which
distinguishes parts of the swampland from the landscape. A simple
observation is that ``gravity is the weakest force" and an intrinsic
UV cutoff for the U(1) gauge theory is obtained in four dimensions.
Recently the author of \cite{Huang:2007gk} proposed to study
inflation under a version of the weak gravity conjecture, namely,
\e \Lambda\leq \lambda^{1/2}M_p\label{wgcs}\q for
$\lambda\phi^4$ theory. This conjecture offers a stringent
constraint on inflation model \cite{Huang:2007gk}. Some related
works on the weak gravity conjecture are discussed in
\cite{Kachru:2006em}.

Unfortunately we still cannot pick out a unique vacuum in string
theory and its seems that a huge number of meta-stable vacua in the
string landscape are available. Once eternal inflation takes place,
inflation can never end globally. A huge number of vacua in the
landscape will be populated even if we start from a definite vacuum.
The eternal inflation provides a natural arena for applying the anthropic
principle. Different parts of spacetime is characterized by
different effective values of physical constants. In this case, some constants
of nature can only be predicted by anthropic reasoning, and some occur by
pure chance. So the key problem for the eternal inflation is to work
out the measure for it \cite{Garriga:2005av}. If eternal inflation
really happens in the early universe, maybe it is not good news
for us because there is no definite prediction of  string
theory.

One possibility to invalidate the anthropic reasoning is that
eternal inflation itself is in the swampland, not in the landscape.
One example was shown in \cite{Chen:2006hs} where the slow-roll
brane inflation in the warped background is not eternal. The
single-field chaotic inflation is in the swampland, but the assisted
chaotic inflation still survives \cite{Huang:2007gk}. In this short
note, we will investigate whether the assisted chaotic inflation can
be eternal or not.

Assisted inflation was proposed and discussed in detail in
\cite{Liddle:1998jc}. Consider $n$ inflaton fields $\phi_i$ $(
i=1,...,n)$ with the same potential $V(\phi_i)$, and there is no
coupling among different inflaton fields. In the spatial flat FRW
universe, the equations of motion read \m H^2&=&{1\over
3M_p^2}\sum_{i=1}^n\(\half {\dot \phi_i^2}+V(\phi_i)\), \\ \ddot
\phi_i&+&3H\dot \phi_i=-{dV(\phi_i)\over d\phi_i}, \quad i=1,...,n.
\n There is a unique late-time attractor where all the inflatons are
equal: $\phi_1=\phi_2=...=\phi_n\equiv \phi$. It is convenient to
adopt the slow-roll parameters for the assisted inflation as
follows: \e \epsilon_H\equiv -{\dot H\over H^2}, \quad \hbox{and}
\quad \eta_H\equiv -{\ddot H\over 2H\dot H}, \label{swp} \q where
the dot denotes the derivative with respect to $t$. When
$\epsilon_H\ll 1$ and $|\eta_H|\ll 1$, the inflatons slowly roll
down their potential and
the equations of motion are simplified to be \m H^2&=&{nV(\phi)\over 3M_p^2},\\
3H\dot\phi&=&-{dV(\phi)\over d\phi}. \n Using the slow-roll
equations of motion and eq. (\ref{swp}), we find \e
\epsilon_H={1\over n}\epsilon_V={1\over n}{M_p^2\over
2}\({dV/d\phi\over V}\)^2. \q The factor $1/n$ helps to achieve the
slow-roll conditions for the case with steep potential for single
inflaton field if the number of the inflatons is sufficiently large.
One of the advantages of the definition of the slow-roll parameters
(\ref{swp}) is that the amplitude of the density perturbations can
be expressed as \e \delta_H\sim {\delta \rho\over \rho}\sim
{H/M_p\over \sqrt{\epsilon_H}}, \label{denp}\q which is the same as
the result in \cite{Liddle:1998jc}. This formulation is available
for the single-field inflation and the assisted inflation.

Generically the quantum fluctuation of a scalar field in de Sitter
space can be modeled by a Gaussian random walk with step size \e
\delta \phi={H\over 2\pi}\q in one Hubble time $H^{-1}$
\cite{Linde:2005ht}. During the same epoch, the classical variation
of one inflaton field is \e \Delta \phi=|\dot \phi|\cdot H^{-1}=
{|dV/d\phi|\over 3H^2}\sim \sqrt{\epsilon_H\over n}M_p. \q Now the
amplitude of the density perturbations (\ref{denp}) becomes \e
\delta_H\sim {1\over \sqrt{n}}{\delta \phi\over \Delta \phi}. \q
Eternal inflation happens when the quantum fluctuation of the energy
density $\delta [\sum_{i=1}^n V(\phi_i)]$ dominates over its
classical variation $\Delta [\sum_{i=1}^n V(\phi_i)]$. Since the
quantum fluctuation of each inflation has uncorrelated Gaussian
distribution, the quantum fluctuation of the energy density takes
the form $\delta [\sum_{i=1}^n V(\phi_i)]=\sqrt{n}(\partial
V/\partial\phi)\delta\phi$. The classical variation of the energy
density is the sum of each inflaton's contribution: $\Delta
[\sum_{i=1}^n V(\phi_i)]=n(\partial V/\partial\phi)\Delta\phi$. So
the eternal assisted inflation happens when \e {\delta \phi \over
\sqrt{n}}> \Delta \phi, \label{aet}\q which implies the amplitude of
the density perturbations is greater than 1.

In slow-roll inflation the Hubble parameter $H$ is roughly a
constant and the number of the degrees of freedom inside Hubble
horizon is bounded by de Sitter entropy \cite{Gibbons:1977mu} \e
S={Area \over 4G}={\pi H^{-2}\over G}. \label{ent}\q
The rate of the change of entropy per e-folding is given by \e
{dS\over dN}={dS\over Hdt}\sim {-\dot H\over GH^4}\sim
\delta_H^{-2}. \q Integrating over the above equation, we get \e
N_{tot}=\int dN\sim \int \delta_H^2dS\leq \int dS=\Delta S, \q where
the total number of non-eternal inflationary e-folds $N_{tot}$ is
bounded by the variation of the entropy during the inflation. Here
we use $\delta_H<1$ for the non-eternal inflation. This result is
valid not only for the single-field inflation, but also for the
assisted inflation. In \cite{ArkaniHamed:2007ky} the authors
conclude that the total number of e-folds for the single-field
inflation is bounded by the de Sitter entropy at the end of
inflation $S_{end}$. Here it is worthy to emphasize  that in fact
the bound on the total number of e-folds is the variation of the
entropy during the inflation, which is more stringent than
\cite{ArkaniHamed:2007ky}. For the large field inflation, the
inflationary entropy at the beginning is much smaller than that at
the end, and then $\Delta S=S_{end}-S_{initial}\sim S_{end}$. But
$\Delta S\ll S_{end}$ for the small field eternal inflation.

From now on we focus on a concrete inflation model: chaotic
inflation with potential \e V(\phi_i)=\half m^2\phi_i^2+{1\over
4}\lambda \phi_i^4. \q The weak gravity conjecture says that the
gravity is the weakest force. When $\lambda\rightarrow 0$, the
intrinsic UV cutoff in (\ref{wgcs}) goes to zero and this field
theory breaks down. First we consider the case in which the
potential is dominated by $\lambda \phi^4$. For $\lambda \phi^4$
assisted inflation, the condition of the eternal inflation
(\ref{aet}) reads \e {H\over \sqrt{n}}>{\lambda \phi^3\over H^2},
\quad \hbox{or} \quad \phi>\phi_E=n^{-1/3}\lambda^{-1/6}M_p \q where
$H\sim \sqrt{n\lambda}\phi^2/M_p$.

We also know that the amplitude of the quantum fluctuations for the
inflaton is ${H/ 2\pi}$ on the Hubble scale $H^{-1}$
\cite{Linde:1993xx} and thus each inflaton field fluctuation
provides a gradient energy density $(\p_\mu \delta \phi)^2\sim
(\delta \phi/H^{-1})^2\sim H^4$. Therefore the total gradient energy
density of the fluctuations of the inflatons is \e n(\p_\mu \delta
\phi)^2\sim nH^4\sim
\frac{n}{M_p^4}\left(\sum_{i=1}^nV(\phi_i)\right)^2 \q which should
be smaller than the total energy density of the inflatons
$\sum_{i=1}^nV(\phi_i)$; otherwise, the inflation is shut off. So we
have $n\sum_{i=1}^nV(\phi_i)=n^2\lambda\phi^4/4<M_p^4$, or
equivalently \e \phi<\phi_G=n^{-1/2}\lambda^{-1/4}M_p. \q The
eternal assisted chaotic inflation happens only when \e\phi_E\leq
\phi_G, \quad \hbox{or}, \quad n<\lambda^{-1/2}. \label{getf}\q For
the perturbative $\lambda\phi^4$ theory, the coupling constant
$\lambda$ is smaller than 1 and $n<\lambda^{-1/2}$ can be easily
satisfied.

On the other hand, there is also a constraint from the weak gravity
conjecture \cite{Huang:2007gk}. Recalling the arguments in
\cite{Huang:2007gk}, the Hubble parameter is the IR cutoff for the
effective field theory and the UV cutoff (\ref{wgcs}) should be
larger than the IR cutoff: \e {n\lambda\phi^4\over M_p^2}\leq
H^2\leq \Lambda^2\sim \lambda M_p^2, \q or equivalently \e
\phi<\phi_W=n^{-1/4}M_p.\label{wgf}\q This result is also valid when
the mass term $\half m^2\phi_i^2$ dominates. For the single-field
chaotic inflation $(n=1)$, $\phi<M_p$ and the slow-roll parameter
$\epsilon_H={M_p^2\over 2}\({dV/d\phi\over V}\)^2={8M_p^2\over
\phi^2}>1$, implying that the single-field chaotic inflation
cannot be achieved in the landscape. For the assisted chaotic
inflation, $\epsilon_H={8M_p^2\over n\phi^2}>{8\over \sqrt{n}}$
which can be smaller than 1 if the number of the inflaton fields is
sufficiently large. That is why the assisted chaotic inflation still
survives. Weak gravity conjecture brings a constraint on the eternal
assisted inflation as follows  \e \phi_E<\phi_W, \quad \hbox{or},
\quad n>\lambda^{-2}. \label{wetf} \q The inequalities in
(\ref{getf}) and (\ref{wetf}) are compatible with each other only
when $\lambda>1$ and $n<1$ which contradicts with our starting
point. The constraint coming from the weak gravity conjecture is
really stringent and even the assisted $\lambda\phi^4$ chaotic
inflation cannot be eternal.

Here it is also interesting for us to consider the constraint from
the observed CMB data. Taking into account the amplitude of the
scalar perturbation spectrum, there is only one free parameter. The
amplitude of the density perturbations $\delta_H$ corresponding to
CMB is given by \e \delta_{H,CMB}={2\over 5}{1\over \sqrt{n}}{\delta
\phi\over \Delta \phi}={n\lambda^\half \phi_{60}^3 \over 2^3\cdot
\sqrt{3}\cdot 5\pi \cdot M_p^3} \label{dhc}\q where $\phi_{60}$ is
the value of the inflaton evaluated at 60 e-folds before the end of
inflation. The number of e-folds $N$ before the end of inflation is
\e N=\int_{t_N}^{t_{end}} Hdt=\int_{\phi_N}^{\phi_{end}} {H\over
\dot \phi}d\phi={n\over M_p^2}\int_{\phi_{end}}^{\phi_N}{V\over
dV/d\phi}d\phi\simeq {n\over 8M_p^2}\phi_N^2, \q or equivalently
$\phi_N=\sqrt{8N/n}M_p$. Now eq. (\ref{dhc}) becomes \e
\delta_{H,CMB}=48\sqrt{\lambda \over n}, \quad \hbox{or}, \quad
\lambda=4.3\times 10^{-4}\delta_{H,CMB}^2n. \label{cb}
\q¡¡Substituting the above constraint into eq. (\ref{getf}), we have
\e n<13\cdot \delta_{H,CMB}^{-2/3}. \q The COBE normalization is
$\delta_{H,CMB}=1.9\times 10^{-5}$ and thus \e n<1.9\times
10^4.\label{dataeg} \q An upper bound on the number of the inflaton
fields is obtained for the eternal $\lambda\phi^4$ inflation if the
gradient energy density is not significant to violate inflation. On
the other hand, if we take into consideration the constraint coming
from the weak gravity conjecture (\ref{wetf}), we find \e
n>1.8\times 10^2\cdot \delta_{H,CMB}^{-4/3}=3.5\times
10^8.\label{dataew} \q We see that the above two inequalities
(\ref{dataeg}) and (\ref{dataew}) cannot be satisfied at the same
time.\footnote{We have ignored some numerical coefficients in the
equations (\ref{getf}), (\ref{wetf}), (\ref{dataeg}) and
(\ref{dataew}). If we recover these coefficients, the contradictions
become more serious and eternal inflation is impossible.}

We also briefly consider the case with the potential dominated by
the mass term. One possible realization of this inflation model in
string theory is proposed in \cite{Dimopoulos:2005ac}. Now the
Hubble parameter is given by $H=\sqrt{n}m\phi/M_p$ and the condition
for the eternal inflation becomes \e \phi>\phi_E=n^{-1/2}\({M_p\over
m}\)^{1/2}M_p. \q Requiring that the gradient energy density is
smaller than the total energy density of the inflatons yields
$n\sum_{i=1}^{n}m^2\phi_i^2/2=n^2m^2\phi^2/2<M_p^4$, or \e
\phi<\phi_G=n^{-1}{M_p^2\over m}.\q The eternal inflation happens
only if \e \phi_E<\phi_G, \quad \hbox{or}, \quad n<{M_p\over m}.
\label{meg}\q Combining with the constraint for the eternal
inflation from the weak gravity conjecture (\ref{wgf}), we have
$\phi_E<\phi_W$ which implies a lower bound on the number of the
inflatons \e n>\({M_p\over m}\)^2.\label{mew}\q Naturally the mass
of the inflaton field should be smaller than the Planck scale and
then eq. (\ref{meg}) contradicts with eq. (\ref{mew}). In this case,
the assisted chaotic inflation is not eternal as well.

To summarize, the chaotic inflation cannot be eternal if we take the
weak gravity conjecture into account; or equivalently, the eternal
chaotic inflation is in the swampland. Here we want to remind the
reader that the weak gravity conjecture \cite{Huang:2007gk} is still
a conjecture and its validity is uncertain. Usually people rely on
the eternal inflation to provide a natural mechanism to populate the
vast number of the meta-stable vacua in the string landscape. If the
eternal inflation really happened, it seems that the anthropic
selection is called for and we will lose the power to make
characteristic predictions from the string theory. However so far
there is not an eternal inflation model really realized in string
theory. We believe that the reason for the absence of the
predictions of the string theory is that we do not understand
string theory well enough. Some key ideas are missing. Maybe the
weak gravity conjecture is just one of those key ideas.

\vspace{.5cm}

\noindent {\bf Acknowledgments}

We would like to thank Yi-Fu Cai for useful discussions.

\end{document}